\documentclass[authoryear,12pt]{elsarticle}
\usepackage[margin=0.80in]{geometry}
\usepackage{float}
\usepackage{adjustbox}
\usepackage{wrapfig}

\usepackage{graphics}
\usepackage[dvips]{color}
\usepackage{graphicx}

\usepackage[font=small,skip=0pt]{caption}
\usepackage{subcaption}

\usepackage{amssymb,amstext,amsmath}
\usepackage{float}
\usepackage{booktabs}
\usepackage{array,ragged2e}
\usepackage[table]{xcolor}
\usepackage{multirow}
\usepackage{algorithm}
\usepackage{algpseudocode}
\usepackage{pifont}
\newcommand{\squeezeup}{\vspace{-10mm}}

\journal{}
\begin{document}
\begin{frontmatter}

\title{Cycle-to-Cycle Queue Length Estimation from Connected Vehicles with Filtering on Primary Parameters}

\author[MCSaddress]{Gurcan Comert\corref{cor1}}
\cortext[cor1]{Corresponding author}
\ead{gurcan.comert@benedict.edu}
\author[MCSaddress]{Negash Begashaw}
\address[MCSaddress]{Computer Sc., Phy., and Engineering Department, Benedict College, 1600 Harden St., Columbia, SC USA 29204}
\begin{abstract}
Estimation models from connected vehicles often assume low level parameters such as arrival rates and market penetration rates as known or estimate them in real-time. At low market penetration rates, such parameter estimators produce large errors making estimated queue lengths inefficient for control or operations applications. In order to improve accuracy of low level parameter estimations, this study investigates the impact of connected vehicles information filtering on queue length estimation models. Filters are used as multilevel real-time estimators. Accuracy is tested against known arrival rate and market penetration rate scenarios using microsimulations. To understand the effectiveness for short-term or for dynamic processes, arrival rates, and market penetration rates are changed every 15 minutes. The results show that with Kalman and Particle filters, parameter estimators are able to find the true values within 15 minutes and meet and surpass the accuracy of known parameter scenarios especially for low market penetration rates. In addition, using last known estimated queue lengths when no connected vehicle is present performs better than inputting average estimated values. Moreover, the study shows that both filtering algorithms are suitable for real-time applications that require less than $0.1$ second computational time.
\end{abstract}
\begin{keyword}Queue length estimation, Kalman filter, Particle filter, connected vehicles, market penetration rate.
\end{keyword}

\end{frontmatter}

\section{Introduction}
Queue length estimation (QLE) with sensor technology has been getting more attention as real-time estimation is becoming feasible.
Queue lengths can be utilized especially in intelligent transportation system operations and control. In this usage of queue lengths,  market penetration levels, coverage of sensors, and accuracy are crucial points to consider.
Given low penetration levels, researchers aim to improve estimators by filtering (\cite{tiaprasertqueue,yin2018kalman}). Real-time performance and simplicity would facilitate methods to be used in simpler low cost controllers. Change in low level parameters is also a concern as known constant parameters would generate inaccurate estimates and very high or low values for low market penetration rates (MPR). Researchers have used filtering to reduce error in such cases. Filtering can be applied in different layers depending on the objective. Since true queue lengths are not known, assuming relatively short unchanging parameters would help to improve estimators. The challenges are multimodality over longer term and underlying unknown distribution for parameters. 
In this paper, we utilize Kalman and Sequential Monte Carlo (e.g., bootstrap, particle) filters to estimate primary parameters more accurately and thereby improve  cycle-to-cycle queue length estimation at traffic intersections within the framework of connected vehicles (CV).



Readers are referred to the survey papers by Zhao et al. (\cite{zhao2019estimation}) and Guo et al.(\cite{guo2019urban}) for general queue length estimation from connected vehicles at traffic intersections, and Asanjaranti et al. (\cite{asanjarani2017parameter}) for queue length estimation. Recent similar studies include methods reconstructing trajectories (\cite{xu2017queue, rompis2018probe,zhao2019various}), cycle-to-cycle estimations (\cite{tan2019cycle}), and real-time applications with high market penetration rates (\cite{mei2019bayesian}). Based on shockwaves and queue dynamics, Yin et al. showed effectiveness of Kalman filter in QLE,  especially within $20 \%$ MPR (\cite{yin2018kalman}). Discrete Fourier transform based filtering was applied by \cite{tiaprasertqueue} on queue length estimation but not on primary parameters. Any overestimation and underestimation on primary parameters would be passed to QLE accuracy. Queue sizes, especially medium to high volume-to-capacity ratios can fluctuate significantly. Filtering estimation QLs would result as low or high trimmed values. Thus, knowing relatively fixed parameters at a level can be better target to filter.  

Stochastic models of QLEs from CVs mainly concerns with stationary (behavior after sufficiently long time interval) or non-stationary (short-time intervals, by cycle, and red phase). Such classification is impacted by treatment of arrival rate, distribution, and market penetration rate (MPR) (probe percentage). For instance, Yang and Menendez showed that 20\% to 30\% MPR is required to control signals  (\cite{yang2018queue}).
Although $20$ to $30\%$ MPR is shown to be required to control signals (\cite{yang2018queue}), in another study Argoteet al. showed different accuracy levels, $\pm10 \%$ with $80 \%$ MPR (\cite{argote2011estimation}). Using sensor fusion with the existing deployed traffic sensing technology (\textit{e.g.}, inductive loops, license plate recognition) was also investigated by the researchers (\cite{wu2013real,li2013estimating,comert2013effect,tan2020fuzing,mandal2020measuring}).  

Even though, higher penetration levels are expected, efficient estimators performing at low MPRs would be desirable as better data management will be required. The question of how much data one needs is important not to overload systems for safety and security reasons. Partially supporting these, in an interesting report (\cite{university2016multi}), researchers have found in a simulation environment that over 200 vehicles at $100 \%$ penetration level signal control from CVs becomes difficult due to computational overhead. Rural roads and isolated intersections would also have lower MPRs. Regardless, increasing QLEs and filtering inconsistent extreme over and under estimations are very important for traffic operations and any partially observed systems. 
\subsection{Contributions of this study}
Simple expectation-based QLE models and estimators with known characteristics in \cite{comert2016queue} enable application of filtering algorithms to deal with measurement of low MPR related errors. The estimation problem where we only assume partially observed system becomes interesting. This approach can be adopted for any similar problem like autonomous vehicles. Studies in estimation of primary parameters using only fundamental CV information (\textit{i.e.}, time, location, speed, and vehicle type etc.) are getting more attention (\cite{comert2016queue,van2018estimation,zhao2019estimation}). But, there are limitations for connected vehicles with low MPRs.
In this study, we aim to improve accuracy at low MPRs by improving the low level parameter estimators 
using filtering algorithms. The resulting formulae would not depend on any primary parameters such as arrival rate or MPR. So, the method can be used for time dependent arrival rates or MPR or under dynamic conditions. 

\section{Problem Definition}
\label{sec_prob}
Estimation of queue lengths is one of the inherent key capabilities of connected and autonomous vehicles for operations control. In this study, we aim to estimate the queue lengths at traffic intersections without knowing the true low level parameters as well as true queue lengths.
CVs can share location and time information with a time reference point (e.g., signal timing, red duration), number of queued vehicles is aimed.
\begin{figure}[!htb]
\centering
\includegraphics[width=0.80\linewidth]{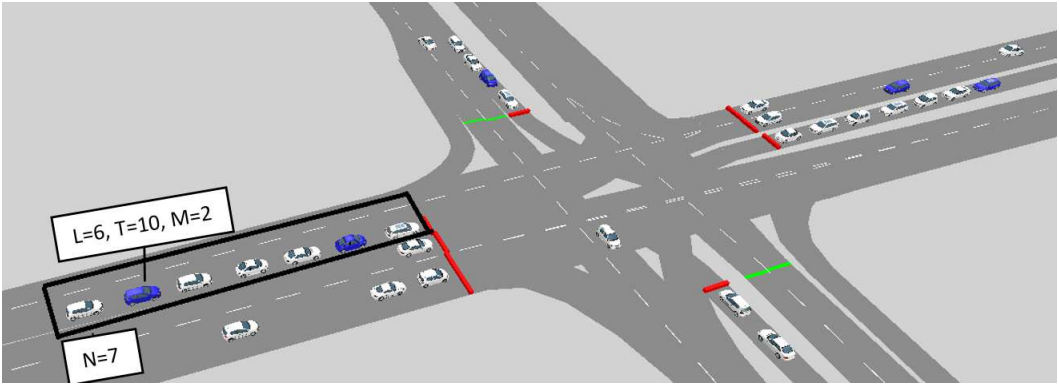}
\caption{An example intersection with multiple approaches generated in Vissim}
\label{fig_int}       
\end{figure}

Considering Fig.~\ref{fig_int} as an example layout, for any approach right before the signal turns green, we aim to estimate the queue length. For instance, consider left turn on east-west bound approach. Assuming that the blue or darker cars are connected vehicles, we try to find total queue length of $N=7$ given last connected vehicle's location $L=6$, number of connected vehicles $M=2$, and arrival time during red $T=10$ seconds $s$ with remaining $R-T$ $s$.


Mathematically, the estimator $E(N_i|L=l_i,T=t_i,M=m_i,R)$ for short target $i$ time intervals is aimed to be described. In the estimator, $N_i$ is cycle-to-cycle total queue lengths at the end of analysis period, $L_i$ location of the last CV, $T_i$ is queue joining time of the CV, and $M_i$ is number of CVs in the queue. Using models in the literature, for Poisson arrivals, simple $l_i+(1-p)\lambda(R-t_i)$ are shown to be an unbiased estimators. Certainly, arrival rate $\lambda$ and market penetration rate $p$ are equally difficult to determine in practice. So, several estimators for $\lambda$ and $p$ are also given in~\cite{comert2016queue}. However, for low market penetration levels less than $30\%$, estimators tend to perform with high variance. Thus, it is intuitive to utilize filters similar to localization idea in robotics. The difference in this problem is that there are no landmarks assumed to correct measurements. 
It is advantageous to assume constant market penetration and arrival rate for relatively short terms, e.g., $15$ to $30$ minutes. This assumption would allow us to locate true values from unimodal distributions. It can be relaxed using desirably nonparametric filtering techniques that would allow multimodal distributions. The nature of the problem can be summarized as follows:
\begin{itemize}
    \item Queue lengths show nonlinear behavior with unknown distribution type.
    \item The underlying process model is also very complex depending on arrival rate, changing signal timing, day of time, and roadway geometry etc.
\end{itemize}
Alternative to such conditions can be listed as discrete Bayes filters and particle filters. Since we are working with constant parameters within short time intervals such as $15$ minutes, we aim to filter parameters for unimodal distribution case. If we are able to get $[\hat{p},\hat{\lambda}]$ values as correctly as possible, then we would be able to estimate QLs more correctly and do not violate the constant parameter within short intervals. In the figures presented in section 4, estimated parameters are given in incremental time interval.
\section{Methodology}
\label{sec_method}
In this study filters are mainly used to estimate primary parameters, arrival rate ($\lambda$) and market penetration level ($p$). The impact of filtering on the lowest level CV time, location, and count information was also checked. Parameter filters would perform better under given consistent behavior of noise levels and described underlying dynamics such as object movement. In this case, no assumption was made for underlying distribution especially for particle filter as it is considered a distribution-free multimodal method. Whereas, KF assumes Gaussian and unimodal underlying process distribution. For both methods, queue length estimation scenarios would be appropriate since unimodal/constant higher level arrival rate and market penetration rates are assumed within 15-minute time intervals.

In order to set up the application,  first consider the estimators in general form of $E(N=n|L=l,T=t,M=m)$=$l+(1-p)\lambda (R-t)$. 
The problem with this estimator is lack of accuracy on $\hat{p}$ and $\hat{\lambda}$, especially at low $p$ ($p \leq 0.30$). We aim to 
 filter and find out true parameter values given short-time intervals \textit{i.e.}, 15 minutes. Estimator combinations that best performed in QLE (\cite{comert2016queue}) were
\begin{eqnarray*}
\hat{p} &= & \left\{ \left(  1 - \frac{m_it_i}{m_it_i + (l_i - m_i)R}  \right) , \frac{m_i}{l_i} \right\} \\
\hat{\lambda} &=& \left\{  \left( \frac{l_i - m_i}{t_i}    + \frac{m_i}{R} \right), \frac{l_i}{t_i} \right\}
\end{eqnarray*}

Thus, queue length estimators are used in Eqs.~(\ref{eqn_est2}) and (\ref{eqn_est3}) for comparison of simple estimator in Eq.~(\ref{eqn_est1}) with $\hat{p}=\frac{m_i}{(l_i/t_i)R}$ and $\hat{\lambda}=l_i/t_i$. Note that this estimator simplifies to 
$\frac{Rl_i}{t_i}-m_i(R-t_i)=R\hat{\lambda}-m_i(R-t_i)$.

\begin{eqnarray}
\label{eqn_est1}
E(\hat{N_i}|l_i,t_i,m_i,q_{i})= \left\{ \begin{array}{ll}
 l_i+\left(1-\frac{m_it_i}{l_iR}\right)\left(\frac{_i}{t_i}\right)(R-t_i), &  l_i>0 \\
E(\hat{N_j}|l_j,t_j,m_j,q_j)\neq0, & j = 1, \cdots, (i-1), l_i = 0 
\end{array}
\right.
\end{eqnarray}
\squeezeup
\begin{eqnarray}
E(\hat{N_i}|l_i,t_i,m_i,q_{i})&=&l+\left(1-\frac{m_i}{l_i}\right)\left(\frac{l_i}{t_i} \right)(R-t) \nonumber \\
&=& l_i+\left(\frac{l_i-m_i}{t_i}\right)(R-t_i)
\label{eqn_est2} 
\end{eqnarray}

\squeezeup
\begin{eqnarray}
E(\hat{N_i}|l_i,t_i,m_i,q_{i})&=&l+\left(1-\frac{m_it_i}{m_it_i+(l_i-m_i)R}\right)\left(\frac{l_i-m_i}{t_i}+ \frac{m_i}{R}\right)(R-t_i) \nonumber \\
&=& m_i+\frac{R(l_i-m_i)}{t_i}
\label{eqn_est3} 
\end{eqnarray}
\subsection{Kalman Filter}
In its simplest form,  Kalman filter (KF) assumes underlying Gaussian measurements or noise (\cite{labbe2014kalman}). KF also performs well if  a physical system is modeled adequately (e.g., vehicle moving dynamics). In the QLE problem, both are not true. We may have different parameter values and no closed-form formulas are available for dynamics of parameters (\textit{i.e.}, $p$, $\lambda$) given connected vehicle information.
 However, it is possible to demonstrate the problem better and can easily be implemented.
 After careful experimentation, observed CV information within $\hat{\lambda}$ as 
$X_i=\min\{(\frac{L_i-M_i}{T_i}+\frac{M_i}{R}),0.272 \}$ 
and
$XC_i=\min\{\frac{XL_i}{XT_i},0.280 \}$ 
was filtered for all ${\lambda,p}$. Similarly, market penetration rate ($\hat{p}$) was estimated and filtered using
$X_i=\min\left\{\left(1-\frac{M_iT_i}{M_iT_i+(L_i-M_i)R}\right),1.00 \right\}$ 
and 
$XC_i=\min\{\frac{XM_i}{XL_i},1.00 \}$.
\begin{algorithm}	
	\caption{Kalman Filter}	
	\begin{algorithmic}
		\State \textbf{assign:} 
		\State $\hat{X_i}=\mu_i$
		\State $S_{X_i}=S_{X_{i-1}}$
		\State \textbf{update:}
		\State $K_{X_i}=S_{X_i}/(S_{X_i}+s)$
		\State $Y_{X_i}=X_i-\hat{X_i}$				
		\State $\mu_{i+1}=\mu_i+K_X Y_X$
		\State $S_{X_{i+1}}=(1-K_{X_i})S_{X_{i}}$
	\end{algorithmic}
	\label{alg1}
\end{algorithm} 

In the algorithm, $S_{X_i}$ is the system uncertainty, $s$ is sensor uncertainty, $K_{X_i}$ is the Kalman gain in cycle $i$, $X_i$ observation in cycle $i$, $Y_{X_i}$ estimation error, and $\mu_{i+1}$ is the state update using Kalman gain and estimation errors.  
\subsection{Particle Filters}
Sequential Monte Carlo (SMC) framework contains many similar filters such as particle, bootstrap, and sequential importance filters with slight changes in the resampling (i.e., update) methods (\cite{liu1998sequential,gustafsson2002particle,doucet2009tutorial, carvalho2010particle,lopes2011particle,bernardo2011particle}). QLE setting has [$p$,$\lambda$] with corresponding weights each showing how likely they represent true $\hat{p}$ and $\hat{\lambda}$ values. 
Unlike usual robotics localization problem landmarks, sensing, or true (or noisy) parameter values are not available in QLE application for low MPRs. Thus, after each movement,  convergence to true parameters (importance resampling) is more challenging. 
Heuristic SMC filtering steps can be written within parameter estimation concepts as follows: 
\begin{itemize}
\item Observe new information from the cycle as triplets [$M$,$L$,$T$] and estimate with [$\hat{\lambda}$, $\hat{p}$].
\item Update the weights of the particles based on the new estimates from the updated observations.
\item Based on the triplet received and new estimation of parameters, resample with replacement to eliminate particles that are highly unlikely. Particles close to the estimates get higher weights.
\item Estimate weighted mean and variance of the filtered $\hat{p}$, $\hat{\lambda}$.
\end{itemize}

Algorithm~\ref{alg2} is adopted from~\cite{lopes2009}.  Initial weights are input as uniform, observations are denoted by $X_{i+1}\sim p(X_{i+1}|Z_{i+1},\lambda)$, and transitions are denoted by $X_{i+1}\sim p(X_{i+1}|X_{i},\lambda)$. Posterior distributions with $p(Z_{i},\lambda|X^i)$ with $X^i=(X_1,...,X_i)$. Since, we do not have movement for particles, a particle filter (Bootstrap filter) with a simpler resampling wheel can be written as propagate: $\{{Z^{(j)}_i}\}_{j=1}^N$ to $\{{\tilde{Z}^{(j)}_{i+1}}\}_{j=1}^N$ via $p(Z_{i+1}|Z^{i})$, generate: $\{{Z^{(j)}_{i+1}}\}_{j=1}^N$ to $\{{\tilde{Z}^{(j)}_{i+1}}\}_{j=1}^N$ via $p(Z_{i+1}|Z^{i})$, and resample with weights: ${{w}^{(j)}_{i+1}}\propto p(X_{i+1}|{\tilde{Z}^{(j)}_{i+1}})$. In fact, without approximately defined dynamics of particles,  full version of particle filter performed very similar with higher computational run time.        
\begin{algorithm}	
	\caption{Particle Filter}	
	\begin{algorithmic}
         
		\State N particles: $X \sim {\mathcal{ N}}(0.0,0.05)$, $\hat{X}_{cycles\times n}$
		\For{\textbf{$cycle$}  \textbf{$i$}}
		\State \textbf{particle} \textbf{jittering:} $X'=X+\epsilon$
		\State $w=f(X_i,X'/(1+{X'}^2,1.00))$				
		\State \textbf{sample:} $n$ \textbf{$from:$} $X'$ \textbf{$with$ \textbf{$p(w)$}}
		\EndFor
		\State $X=X'$, $\hat{X}_{i,1:n}=X$, \textbf{median}(\textbf{$\hat{X}$}) 
		
	\end{algorithmic}
	\label{alg2}
\end{algorithm}

In algorithm~\ref{alg2}, numerical values are selected after trial and errors on one series (e.g., $\rho=0.88$ and $p=0.10$). Low jittering noise was selected to prevent low resampling or fast bad convergence $\epsilon={\mathcal{N}}(0.0,0.0005)$
\section{Numerical Results}
\label{sec_num}
Numerical examples are given to show performance of the filtering approach. Analysis include discussions for both filters and queue length estimation from connected vehicles for fixed and varying primary parameters. Simulated intersection queue data are used in the examples. An isolated signal with fixed signal control in Vissim is used to test the estimators with filtered vehicle information after revising the network generated in \cite{comert2016queue} and default queue vehicle in Vissim is used (\cite{liu2019real}). Simple intersection is used only for generating queueing data. As mentioned earlier estimators can be adopted for approaches with a standing queue.
\begin{figure}[!htb]
\centering
\includegraphics[width=0.95\linewidth]{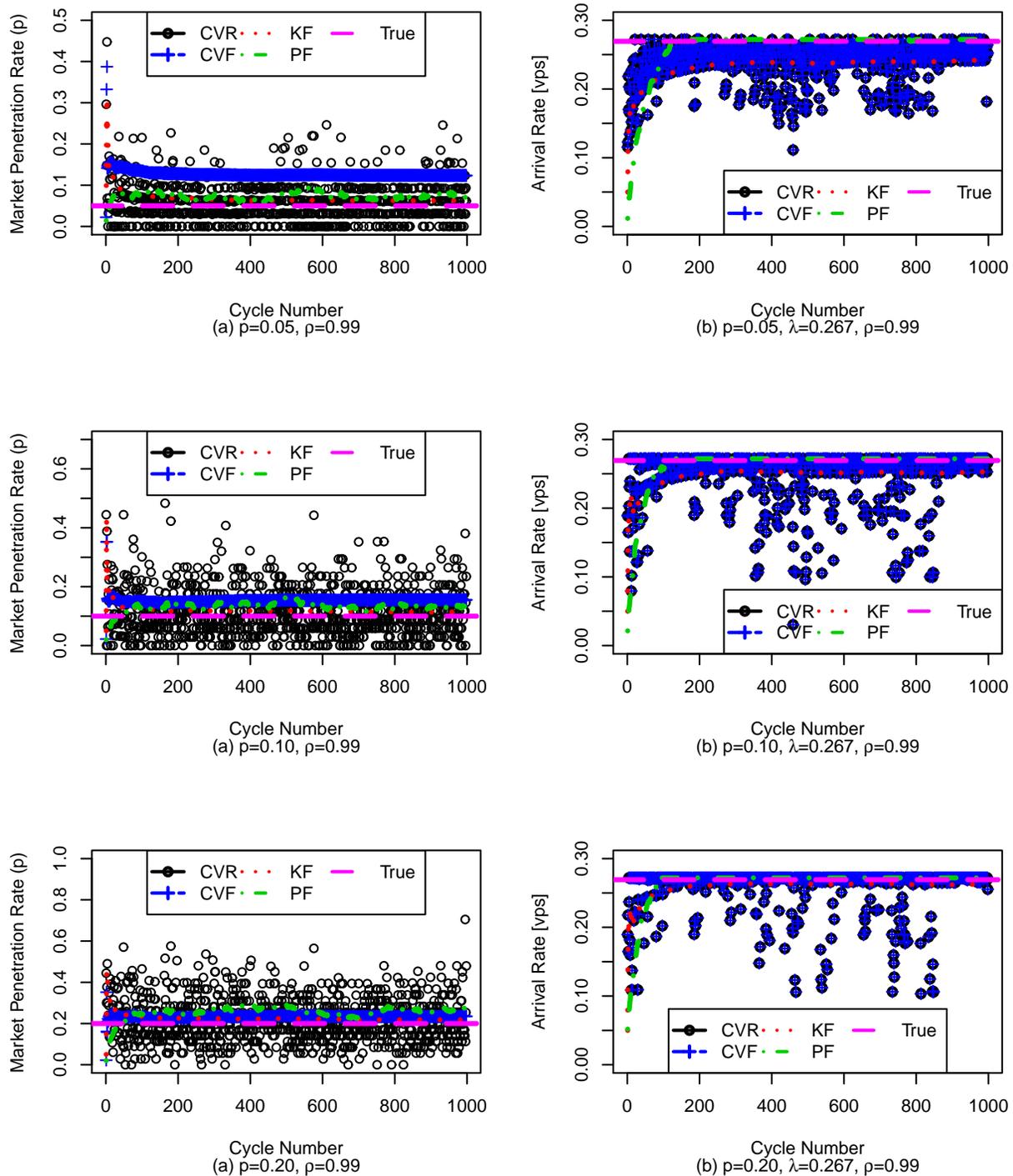}
\caption{Parameter estimation example with filters at $p \%$, $\rho=0.99$ as average of 3 random seeds}
\label{fig_par}
\end{figure}

Note that given required information, estimators can be used under any condition and signal control. Queue lengths that we observe show an autocorrelated times series as dynamic stochastic process.  Thus, if estimators perform well here, they can be used under any conditions as only fundamental vehicle information  location, time, count, and time period of estimation (e.g., signal timing such as red, green duration) was assumed to be known.
\begin{figure}[!htb]
\centering
\includegraphics[width=0.95\linewidth]{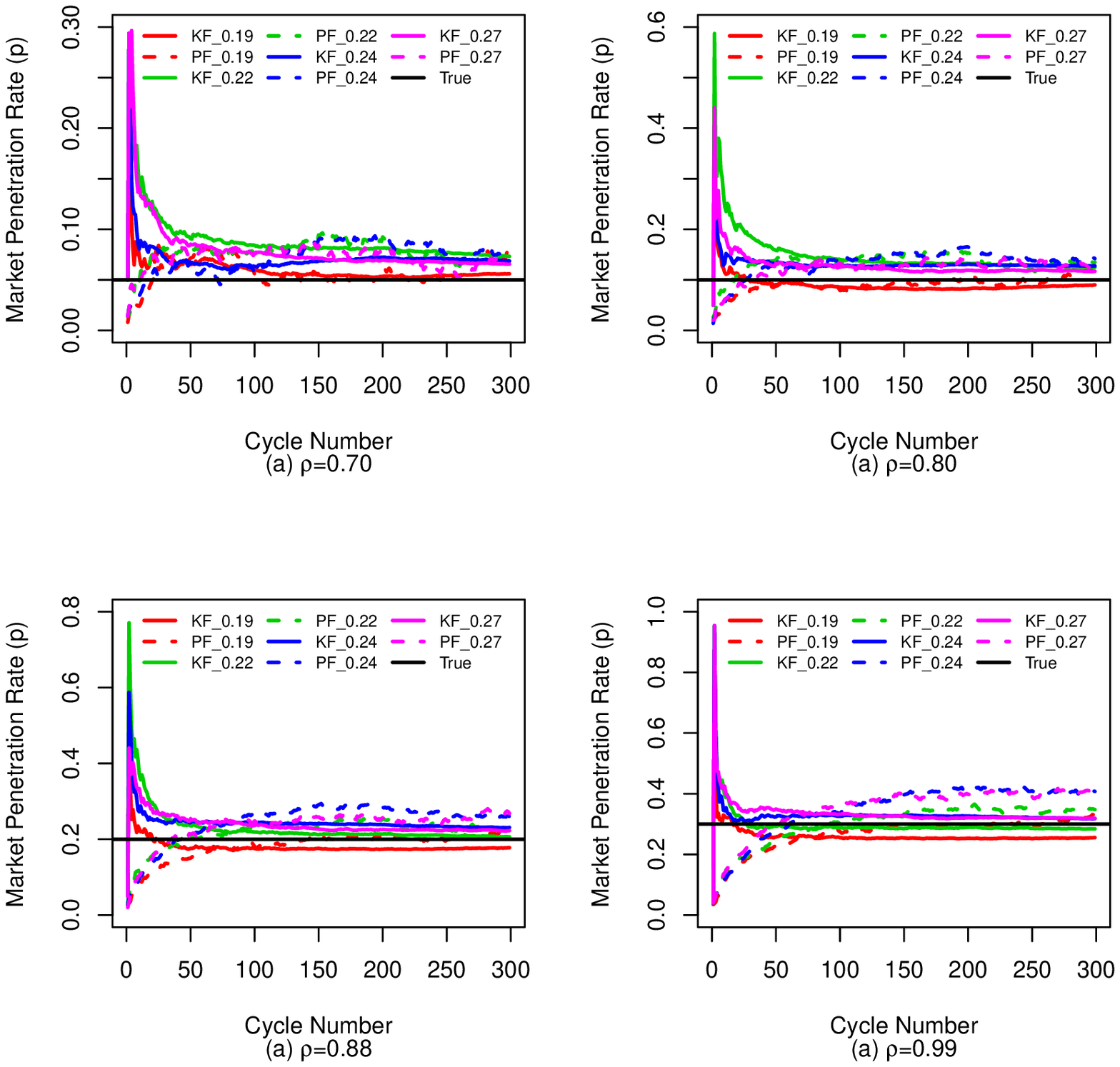}
\caption{Estimation examples for $p$ with filters at $p={0.05,0.10,0.20,0.30}$, $\rho={0.70,0.80,0.89,0.99}$ as average of 3 random seeds}
\label{fig_p}
\end{figure}

\begin{figure}[!htb]
\centering
\includegraphics[width=0.95\linewidth]{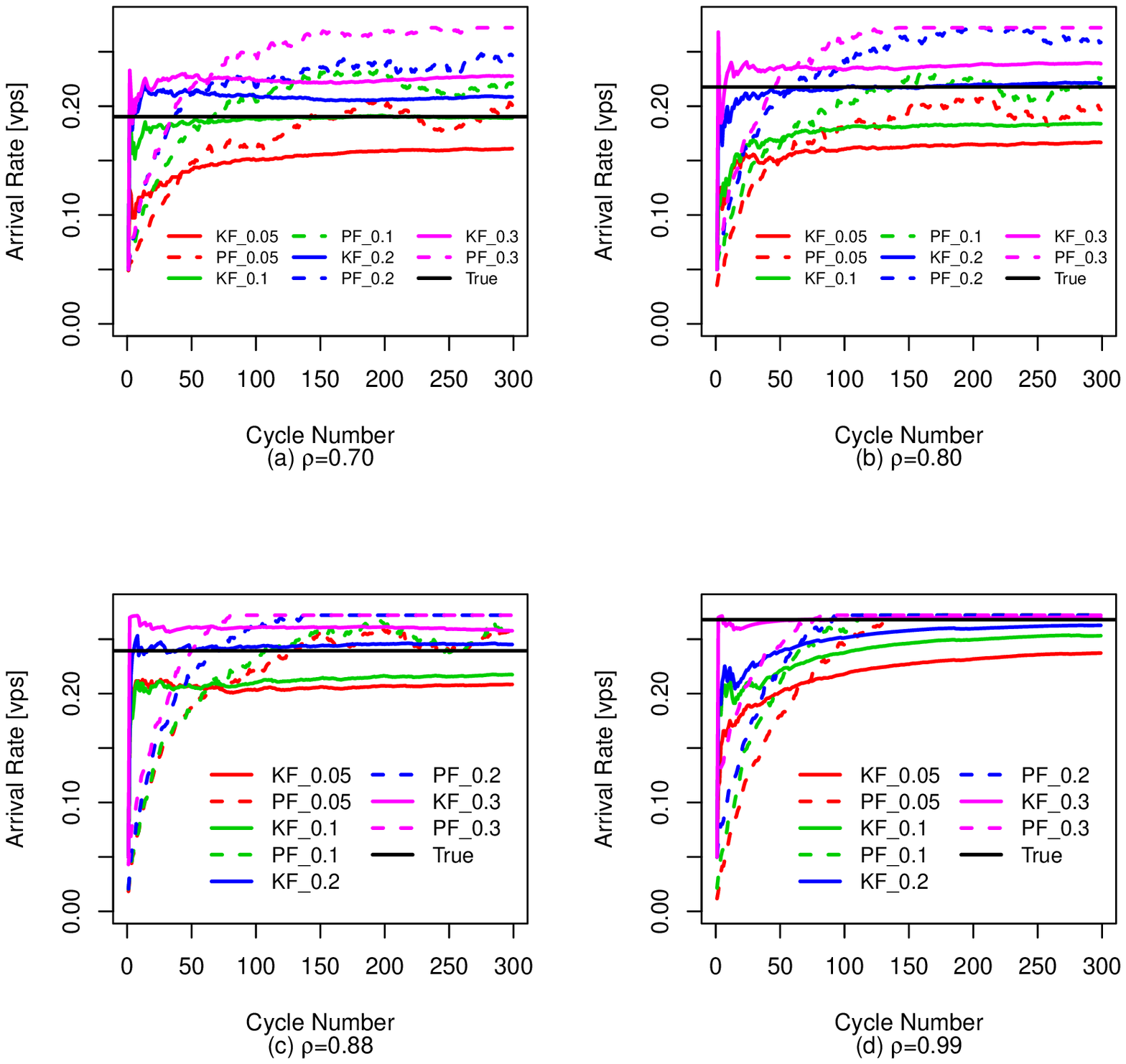}
\caption{$\lambda$ estimation examples with filters at $p={0.05,0.10,0.20,0.30}$, $\rho={0.70,0.80,0.89,0.99}$ as average of 3 random seeds}
\label{fig_lam}
\end{figure}

Vehicle trace files generated from microsimulations are used to evaluate queue length estimation without and with initial queue cases for arrival rates of $\lambda=\{0.163,0.190,0.218,0.239,0.267\}$ vehicle per second ($vps$) that correspond to volume-to-capacity ratios of $\rho=\{0.60,0.70,0.80,0.89,0.99\}$ and $p=\{0.001,0.05,0.10,0.20,...,0.90\}$ for the desired simple intersection with $90$ seconds ($s$) cycle length equal $45$ $s$ red/green split. Note that our approach is QLE given partial information about the system and can be applied for any approach with a standing queue at a certain time period (\textit{e.g.}, red duration).
\subsection{Fixed parameters}
\begin{figure}[!htb]
\centering
\includegraphics[width=0.95\linewidth]{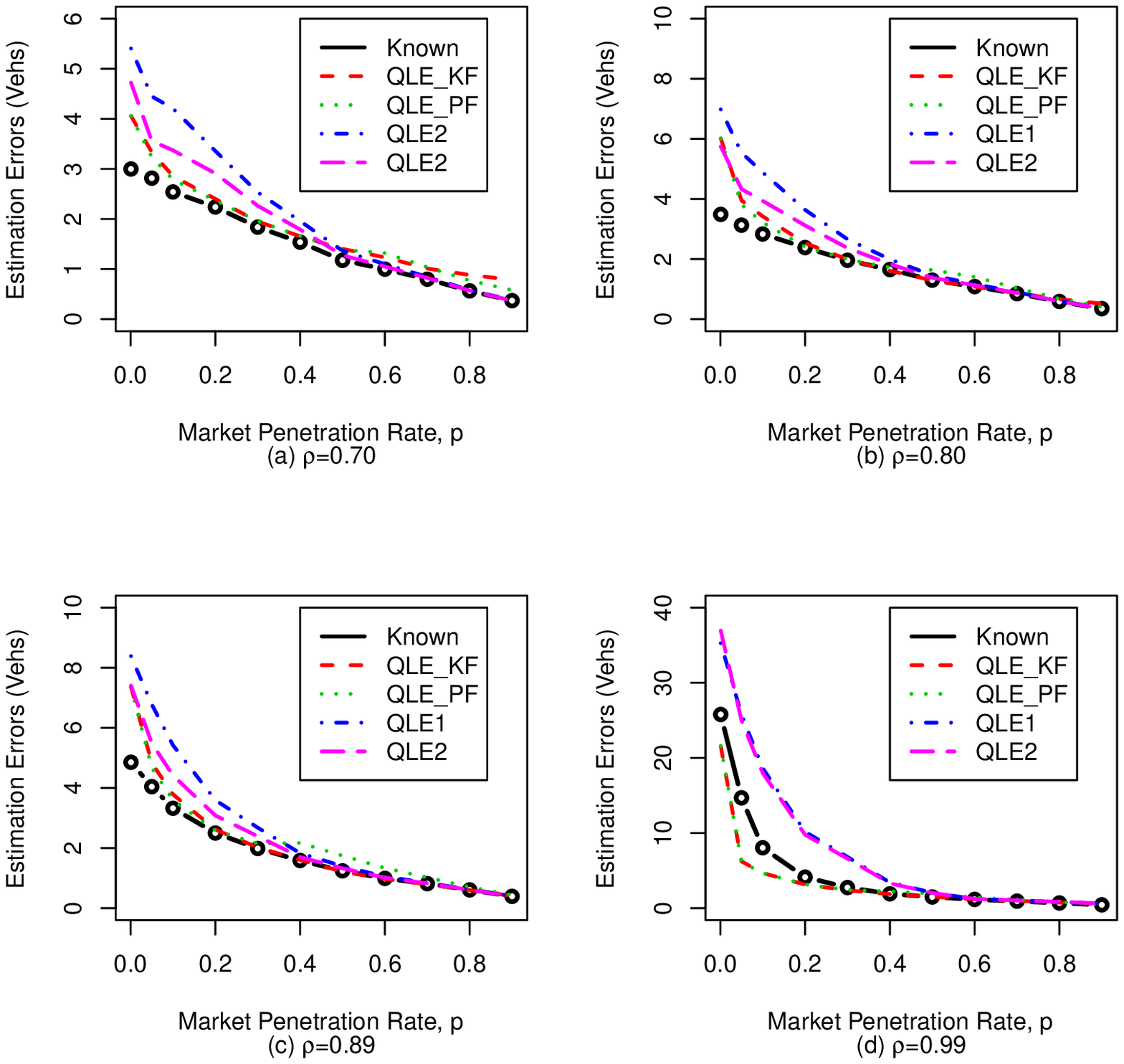}
\caption{Performance with filters for different $p$, $\lambda$, and $\rho$}
\label{fig_vardf}
\end{figure}

\begin{figure}[!htb]
\centering
\includegraphics[width=0.95\linewidth]{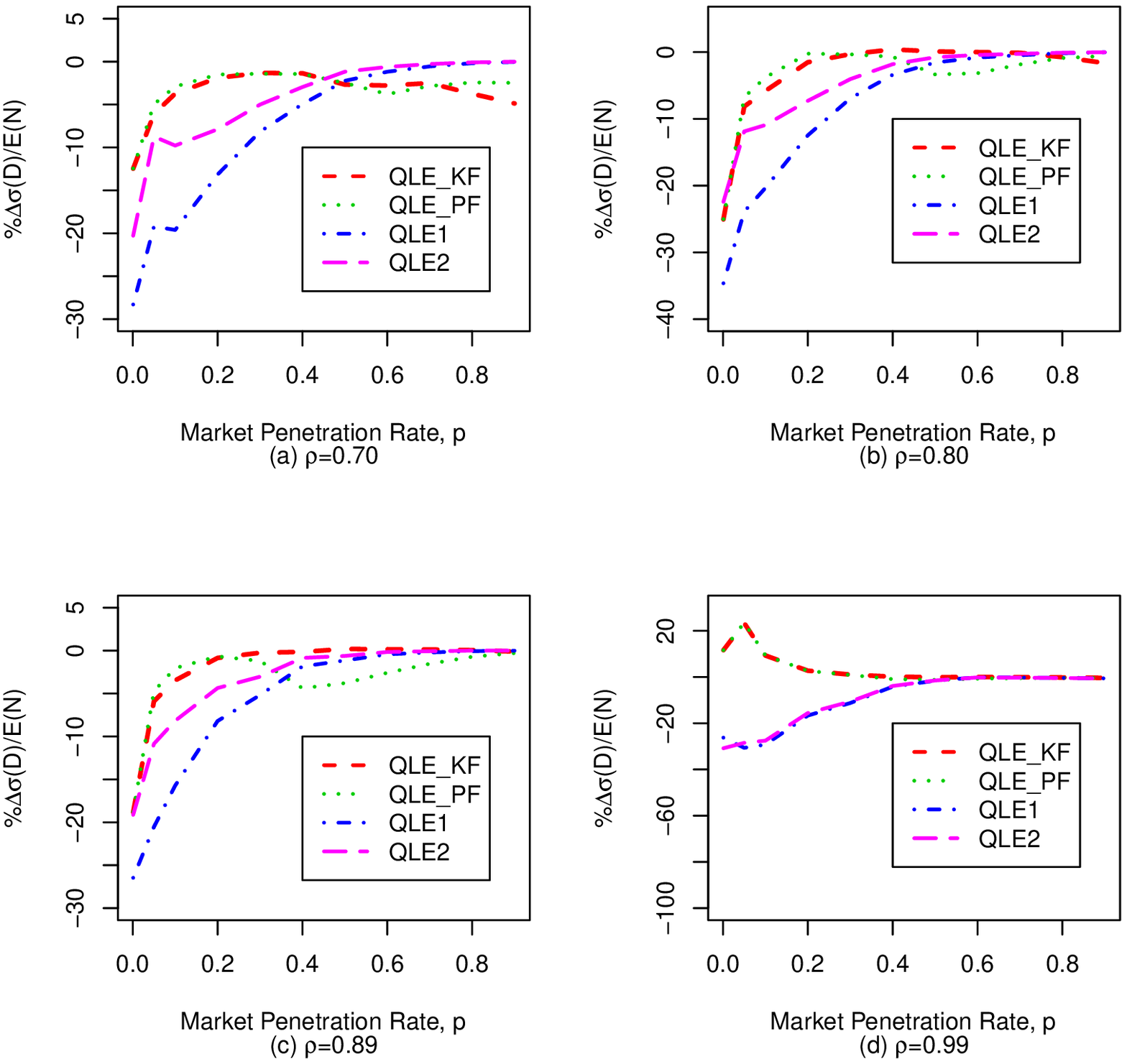}
\caption{Impact of filters in estimation $\%\Delta\sqrt V(D)/E(N)$ for different $p$, $\lambda$, and $\rho$ average of 3 random seeds}
\label{fig_dvardf}
\end{figure}

\begin{figure}[!htb]
\centering
\includegraphics[width=0.95\linewidth]{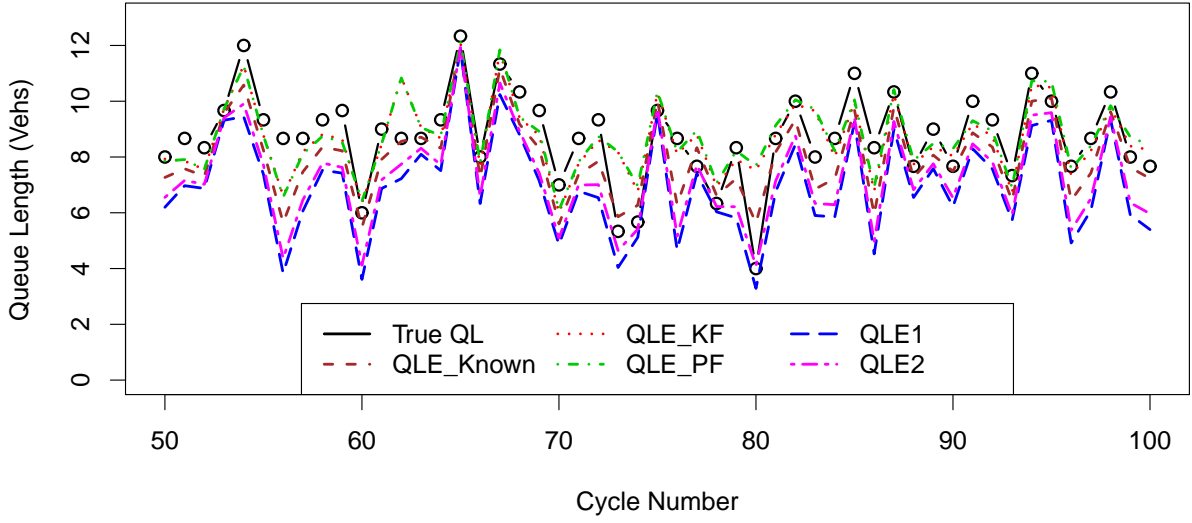}
\caption{Estimation example with filters at $p=5\%$, $\rho=0.99$ as average of 3 random seeds}
\label{fig_acc}
\end{figure}

\begin{figure}[!htb]
\centering
\includegraphics[width=0.95\linewidth]{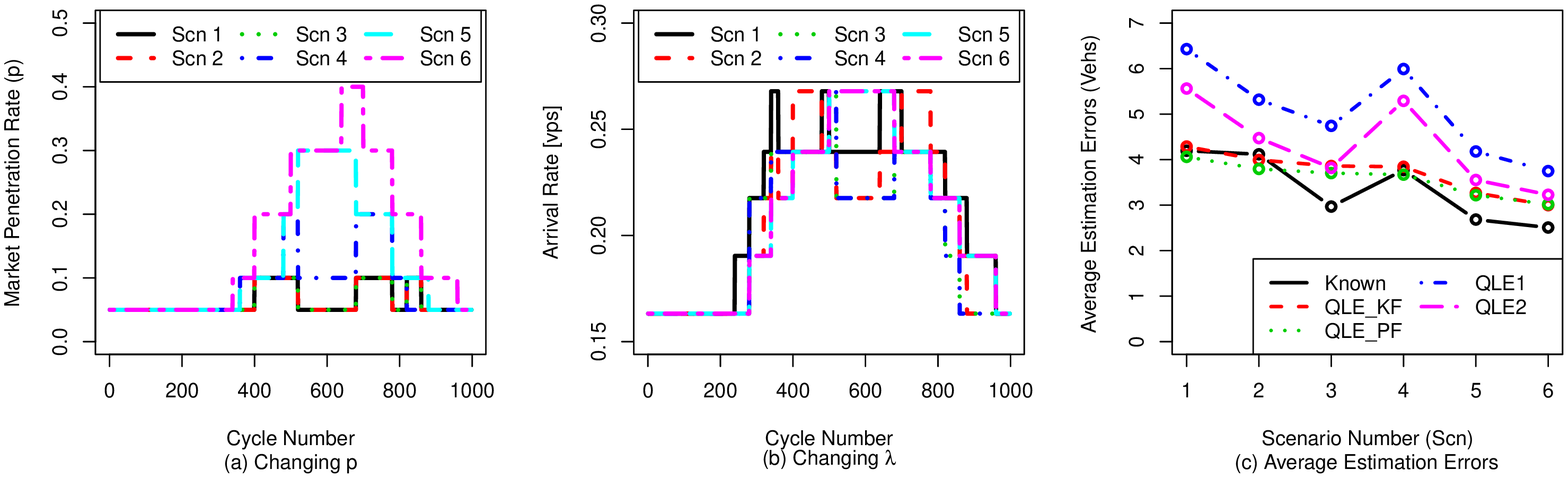}
\caption{Scenarios with changing $p$, $\lambda$, and resulting estimation errors in $\sqrt{V(D)}$ [vehs]}
\label{fig_scn}
\end{figure}

Cycle-to-cycle updated parameter estimators using filtering that do not involve cumulative connected vehicles information are given in Figs.~\ref{fig_par}-\ref{fig_scn}. In order to check the impact in lowest level information (\textit{i.e.}, applied on $L$,$T$,and $M$), estimator values with raw CV are denoted by $CVR$ and the values with filtered CV information are denoted by $CVF$. 

Impact is more seen in $\hat{p}$.  However, resulting QLEs are not much impacted when using raw or filtered CV values.

From Fig.~\ref{fig_par}, it can be seen that highly dispersed ratios are filtered to point desired parameter level. However, we are not able to obtain true parameters for $p<0.20$. At MPR $p=0.20$, parameter estimators are improved 
and with filtering, we are able to track the underlying true parameter values $p$ and $\lambda$ very closely. 

Estimator performances for $\hat{\lambda}$ for different $p$ levels and $\hat{p}$ for different $\lambda$ levels are given in Figs.~\ref{fig_p} and \ref{fig_lam}, respectively. In Fig.~\ref{fig_lam}, KF is able to get closer to true values at low volume-to-capacity ratio $\rho=0.70$ and $p=0.10$. In all cases, as $p$ gets higher, estimator performs better. Similarly, in Fig.~\ref{fig_p}, it can be seen that estimation with filters perform well for all $\lambda$ and $p$ values. Both KF and PF show similar behavior.

In Fig.~\ref{fig_vardf}, estimation errors with and without filtering are presented. Filtering improves the estimation errors compared to unknown primary estimators (denoted by QLE1 and QLE2). With filtering, at $20\%$ market penetration level, errors with known primary parameters can be met. As volume-to-capacity ratio increases, filtering is able to meet and improve the queue length estimation compared to known case. 
\begin{figure}[!htb]
\centering
\includegraphics[width=0.95\linewidth]{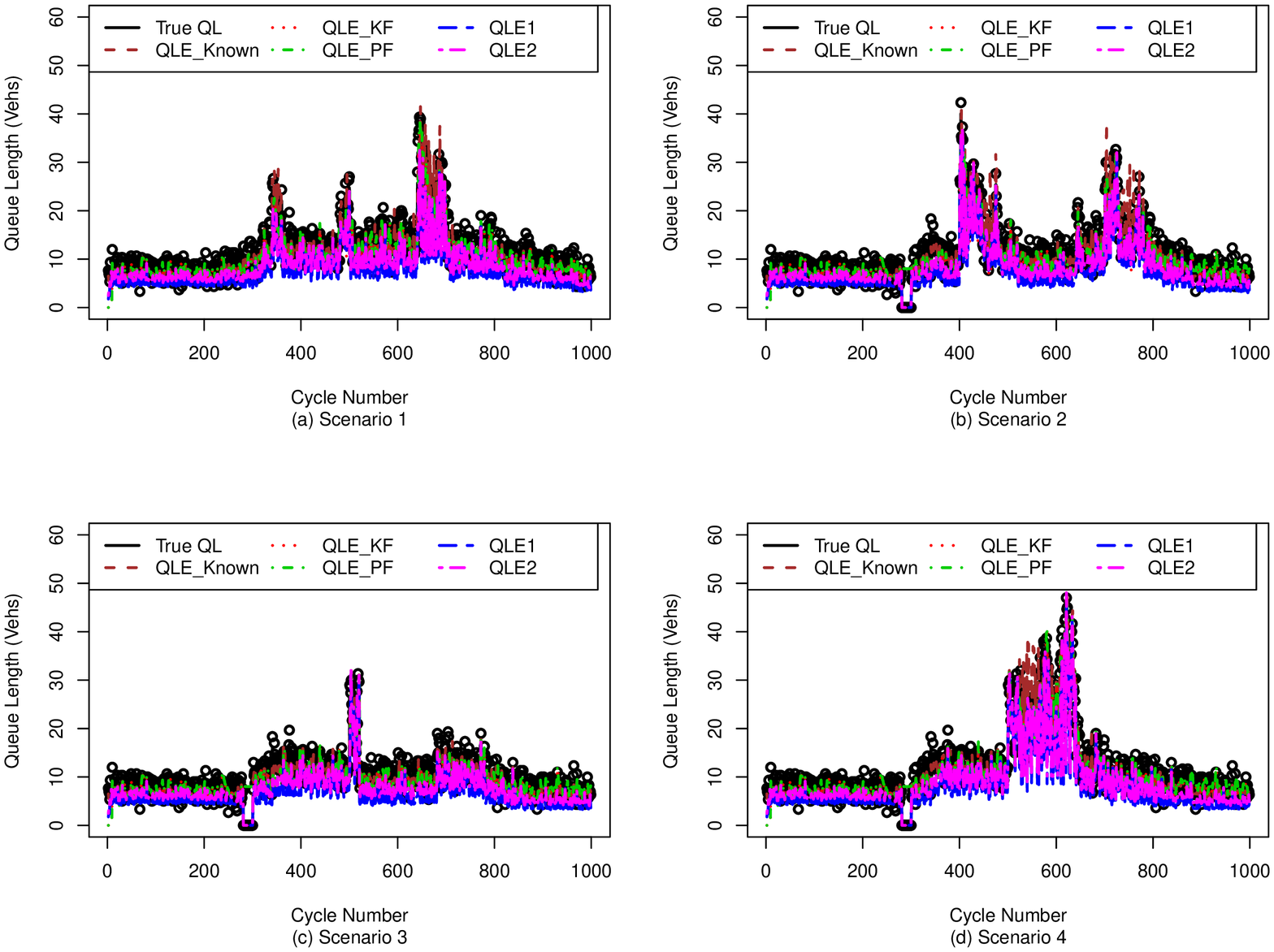}
\caption{Queue length estimation examples}
\label{fig_sc}
\end{figure}

The differences in the form of $\%\Delta\sqrt V(D)/E(N)$ from QLE with known ($p$, $\lambda$) errors can be seen in Figs.~\ref{fig_dvardf}a-d. In all $\rho$ values up to $40 \%$ MPR, QLE with filtering shows improved errors, \textit{i.e.}, lower than QLE1 and QLE2. This is desirable as accuracy improvement is needed at low MPR values. Only for $rho=0.70$, after $50 \%$ MPR filtering shows worse errors. However, these errors are already very low (less than 1 vehicle in Fig.~\ref{fig_vardf}). QLE with filtering shows better performance than QLE with known ($p$, $\lambda$) up to $40 \%$ in $\rho=0.99$.
\begin{figure}[!htb]
\centering
\includegraphics[width=0.95\linewidth]{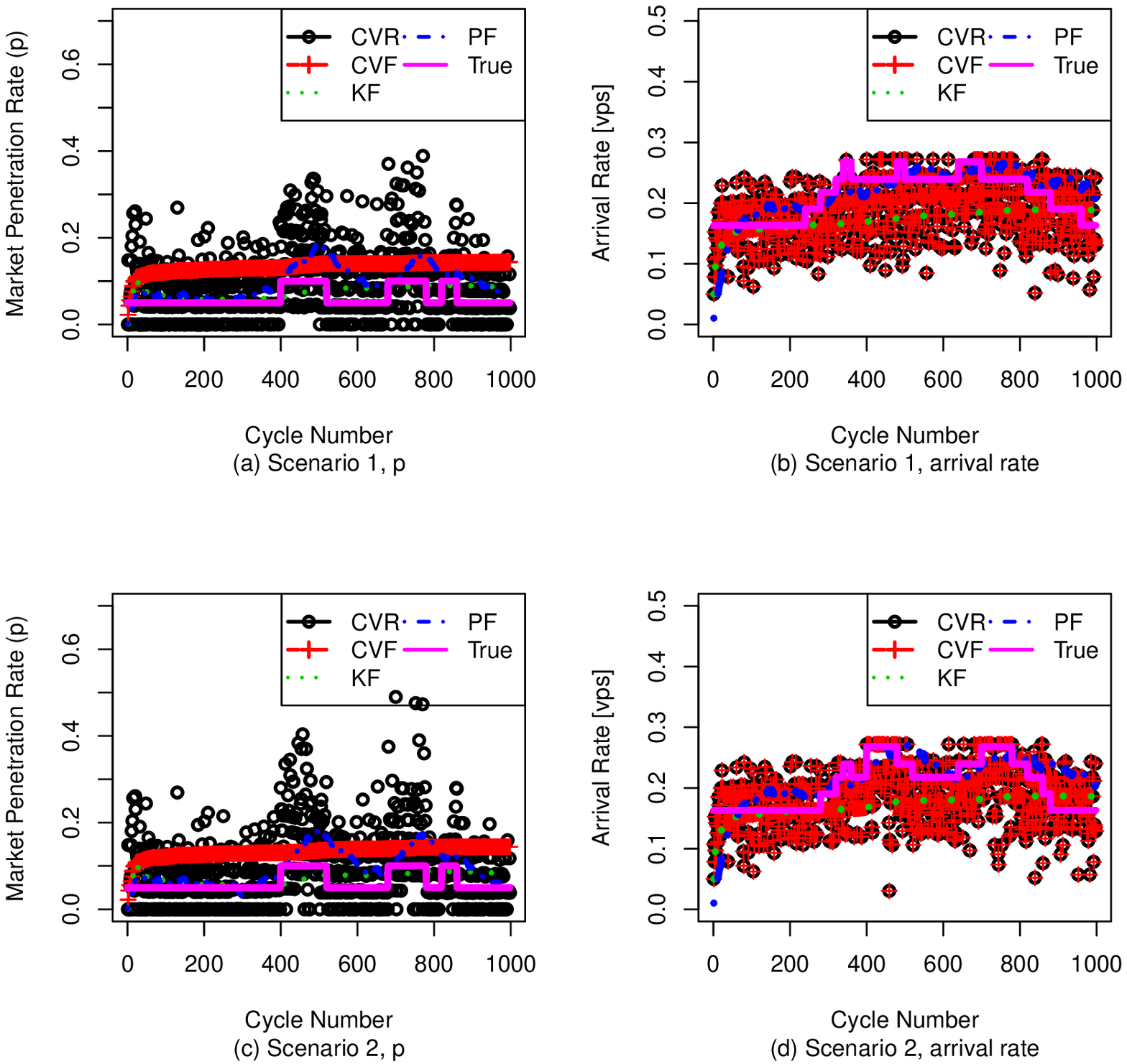}
\caption{Parameter estimation example with filters at $p \%$, $\rho=0.99$ as average of 3 random seeds}
\label{fig_scpar}
\end{figure}

Example of queue length estimation is shown in Fig.~\ref{fig_acc} for $p=5 \%$ and $\rho =0.99$. Filtered estimators are able to closely follow true QLs. Extreme reactions due to misrepresenting CAVs information are filtered. 

All the examples so far have been on single parameter for relatively long time. Estimators are able to perform well in $50$ cycles or $1.25$ hours. Next, we consider changing parameter cases and performance of estimation with filters. 
\subsection{Dynamic parameters}
In this subsection, estimators' performance are shown when parameters change for shorter time intervals. Since the physics of queue evolution is not defined in closed form for KF, it is expected that PF performs better for multi-regime parameters as shown in Fig.~\ref{fig_scn}a-b. These scenarios are arbitrarily generated by increasing and decreasing with a peak to mimic changing traffic conditions. Setting up full factor experiments is beyond the scope of this paper.   

Overall average estimation errors from these scenarios are also given in Fig.~\ref{fig_scn}c. From the figure, we can deduce that PF performs slightly better than KF in all scenarios. Estimation with filtering outperforms unknown parameter cases and closely follows the estimation errors of known ($p$, $\lambda$) QLE. 

Example queue length estimations are given in Fig.~\ref{fig_sc} for 4 scenarios where overall we can see filtering helps to more closely follow true queue lengths.
Behavior of KF and PF on changing parameter estimations are shown in Fig.~\ref{fig_scpar}. PF is expected to perform better under dynamic $p$ and $\lambda$ behaviors. It is also observed that for low $p$ values, using last estimated queue length performs much better than using average values until then. This is due to volatile behavior of queue lengths instead of a level we are able to catch high and low points better with last known point missing data amputation.

For computational times, for a single scenario for $1000$ cycles running KF for $\lambda$ and $p$ estimations on a PC with 8GB of memory, Pentium I5 Quad-Core CPU took $0.03$ $s$ and $0.04$ $s$, respectively. 
For running PF took $3.50$ $s$ and $3.31$ $s$, respectively. As per cycle calculations are well below $0.1$ $s$, both filtering algorithms can be utilized in real-time cycle-to-cycle control and safety critical applications.
\section{Conclusions}
In this study, we investigated the impact of localization framework on queue length estimation at low market penetration rates. We used two different filtering methods:
\begin{description}
\item[(i)] Kalman filter: unimodal and Gaussian noise
\item[(ii)] Particle filter: multimodal and noise from any distribution
\end{description}


The filters are fitted to queue length estimation framework where no approximately true feedback is available.

Two cases of fixed arrival and market penetration rate parameters and changing parameter are analyzed. Queue length estimation with filters show promising results improving estimation by $20 \%$ for high volume-to-capacity ratio of $\rho=0.99$. For lower $\rho$ levels, with filtering by $30  \%$ market penetration rates, known queue length estimation error are matched.

Both estimation models and filters are simple
and enable real-time applications with no significant overhead. For future research, alternative queue length estimation models will be included for more realistic arrival and service time distributions. 
\section*{Acknowledgments}
This study is partially supported by the Center for Connected Multimodal Mobility ($C^{2}M^{2}$) (USDOT Tier 1 University Transportation Center) headquartered at Clemson University, Clemson, South Carolina. Any opinions, findings, and conclusions or recommendations expressed in this material are those of the authors and do not necessarily reflect the views of the Center for Connected Multimodal Mobility ($C^{2}M^{2}$) and the official policy or position of the USDOT/OST-R, or any State or other entity, and the U.S. Government assumes no liability for the contents or use thereof. It is also partially supported by U.S. Department of Homeland Security Summer Research Team Program Follow-On grant and NSF Grants Nos. 1719501, 1436222, 1954532, and 1400991.

\bibliographystyle{elsarticle-harv}
\bibliography{projects}

\end{document}